\documentclass[aps,prd,reprint,amsmath,amssymb,nofootinbib]{revtex4-2}

\usepackage{graphicx}
\usepackage{bm}

\newcommand{\Tp}{T_{\perp}}
\newcommand{\Tpz}{T_{\perp 0}}
\newcommand{\lB}{\ell_B}
\newcommand{\lBz}{\ell_{B0}}
\newcommand{\wc}{\omega_c}
\newcommand{\wcz}{\omega_{c0}}
\newcommand{\coty}{\cot(y_0/\lambda)}
\newcommand{\eV}{\,\mathrm{eV}}
\newcommand{\meV}{\,\mathrm{meV}}
\newcommand{\keV}{\,\mathrm{keV}}

\begin{document}

\title{Quantum limits of adiabatic drift filtering}

\author{Wonyong Chung}
\affiliation{Department of Physics, Princeton University, Princeton, New Jersey 08544, USA}

\date{\today}

\begin{abstract}
The transverse adiabatic drift filter takes a charged particle's transverse kinetic
energy down an exponential magnetic profile --- keV to sub-eV within a meter --- while a
crossed electrostatic field holds the trajectory straight; its published theory is
classical. We give the quantum theory of the instrument and locate its limits. In the
classical foundation, the drift flow conserves an exact invariant, the
adiabatic invariant suffers no secular change at any order of the guiding-center
expansion on or off the balance condition, i.e. compressing the Landau ladder
but never changing the index, and the off-balance trajectory is given in closed
form, confirmed by exact relativistic orbit integration. Quantum mechanically
the drift sector is exactly solvable: near balance the flow is a canonical squeeze, so
the exponential gain magnifies signal and quantum spread by the same factor, and the
resolving power of the dispersive mode is fixed at the entrance plane. For an entrance
state at the coherent scale ($\Delta Y_0 = \lBz/\sqrt{2}$) the result is a
length-independent entrance limit,
$\sigma_{\mathrm{ent}} = (\Tpz \coty/\lambda)\,\lBz/\sqrt{2} \approx
4.8\meV\cdot\coty$ at $1$~T for an $18.6\keV$ electron with $\lambda = 7$~cm --- the
dominant quantum limit of every sub-meV selection window contemplated at tesla-scale
entry. The energy balance itself has a single $\hbar$-controlled noise channel, the shot
noise of spontaneous cyclotron emission, which we evaluate exactly: the mean radiated
energy is independent of the electron's energy across balanced settings at
non-relativistic order, the fluctuation is
$\sigma^2_{\mathrm{rad}} = \tfrac{3}{4}\,\hbar\wcz E_{\mathrm{rad}}$ with thermal and
spectral corrections evaluated --- thermal enhancement is quartically suppressed below
$\hbar\wcz/k_B$, with coefficient $8\pi^4/15$ --- and the floor is identical for
calorimetric and position readouts. The entrance limit falls with entrance field as
$B_0^{-1/2}$; the radiation floor rises as $B_0^{2}$; their quadrature sum has a
closed-form interior optimum --- near $2.5$--$5$~T at the published geometry, with a
universal $4{:}1$ variance partition between entrance limit and radiation floor --- in a
design space where slowing the forward drive and squeezing the entrance guiding-center
state lower the floor and the required field together. Every result reduces to the published classical
claims as $\hbar \to 0$.
\end{abstract}

\maketitle

\section{Introduction}
\label{sec:intro}

An electromagnetic filter that drains transverse kinetic energy adiabatically --- a
magnetic field falling as $e^{-z/\lambda}$ with a crossed electric field matched to it
--- was introduced as the energy-analyzing element of a planned relic-neutrino-capture
measurement~\cite{betti2019design}, and has since been demonstrated as an instrument
class in its own right~\cite{apponi2022implementation,farino2025slowed}: the
start-to-end field ratio is the energy-reduction ratio, delivered in linear distance at
an exponential rate. Where the MAC-E
filter~\cite{beamson1980,lobashev1985,picard1992,katrin2004design} collimates momentum
and applies an integral high-pass in a longitudinal gradient, the drift filter selects a
transverse slice of phase space and magnifies it; the two instruments are orthogonal
configurations of the same adiabatic invariant.

The published theory of the drift filter is classical, and its central resolution
statement --- an acceptance window closing exponentially with depth, with no stated
floor~\cite{chung2025thesis,betti2019design} --- is a classical statement; the
quantum analyses within the parent measurement program have concerned its other
elements~\cite{apponi2022heisenberg}. This paper asks what quantum mechanics permits of
the filter itself. The question has become concrete for two reasons. First, the design
targets have entered the regime where quantum scales are visible: a $10\meV$
transmission window sits within two orders of magnitude of the entry Landau spacing at
tesla fields --- the regime of the one-electron quantum
cyclotron~\cite{peil1999,hanneke2008} --- and proposed refinements reach further down.
Second, the instrument is adjustable --- the balance condition, the exit configuration,
and the geometry are design choices rather than constraints --- so limits derived for
the mechanism, rather than for one apparatus, determine what the class can ever do.

Our results are organized as four theorems and their consequences.

\emph{The invariant theorem (Sec.~\ref{sec:classical}).} The drift flow conserves
exactly the guiding-center energy $D = \mu B(z) + q\phi(y,z)$, and the adiabatic
invariant $\mu$ itself has no secular drift at any order of the guiding-center
expansion, on or off the balance condition. Off-balance operation --- the generalized,
dispersive instrument --- loses nothing in invariant quality: the filter compresses the
Landau spacing but never lowers the index. The closed form for the
off-balance trajectory is confirmed by exact relativistic orbit integration.

\emph{The entrance-limit theorem (Secs.~\ref{sec:quantum}--\ref{sec:ent}).} In canonical
drift variables the near-balance dynamics is a pure squeeze: signal and quantum
uncertainty are magnified by the same $e^{z/\lambda}$, so no filter length improves the
resolution of the dispersive mode. The floor is the entrance guiding-center spread
converted by the entrance gradient --- length-independent, $\hbar$ entering once through
the magnetic length --- and the gyro-orbit smear, the entrance limit, and the local
magnetic length form a strict hierarchy of three floors of which the local one never
dominates.

\emph{The radiation-floor theorem (Sec.~\ref{sec:radiation}).} The only
$\hbar$-controlled noise on the instrument's energy balance is the shot noise of
spontaneous cyclotron emission during the transit. We compute the mean radiated energy
--- an invariance: across energy-retuned balanced settings, at non-relativistic order,
it does not depend on the electron's energy --- its compound-Poisson fluctuation, its
independence of readout mode, its thermal correction, and the noise contribution of any
upstream dwell at full field.

\emph{The optimum theorem (Sec.~\ref{sec:optimum}).} The entrance limit scales as
$B_0^{-1/2}$ and the radiation floor as $B_0^{2}$, so the dispersive instrument has a
closed-form optimal entrance field with a universal $4{:}1$ variance partition between
entrance limit and radiation floor, and a design space whose knobs --- forward-drive
angle, decay length, entrance squeezing, emission-environment engineering --- carry
exact exponents.

Section~\ref{sec:adiabaticlimit} bounds the instrument's length through the adiabatic
limit;
Sec.~\ref{sec:meaning} states what the limits imply for the instrument class and points
to the companion paper on physics reach. Throughout, the instrument's published
configuration ($18.6\keV$ electrons, $B_0 = 1$--$10$~T, $\lambda = 5$--$7$~cm, exit near
$1$~mT) supplies reference numbers, but every statement is for the mechanism class.

\section{The instrument, its exact invariant, and the corrected classical foundation}
\label{sec:classical}

\subsection{Fields and balance}
\label{sec:fields}

The configuration is the harmonic pair
\begin{align}
\phi &= V_0 \sin(y/\lambda)\, e^{-z/\lambda},
\nonumber\\
\mathbf{B} &= B_0\bigl(\cos(x/\lambda),\, 0,\, -\sin(x/\lambda)\bigr) e^{-z/\lambda},
\label{eq:fields}
\end{align}
a vacuum solution with $|\mathbf{B}| = B_0 e^{-z/\lambda}$ everywhere. On the invariant
plane $x = 0$ (exactly preserved by the dynamics), the $E\times B$ drift components are
$z$-independent --- the shared decay length cancels between the fields --- and the
gradient-$B$ drift is constant in magnitude, $\mu/(e\lambda)$. A straight trajectory
requires the vertical components to cancel:
\begin{equation}
e V_0 \sin(y_0/\lambda) = \mu B_0
\;\;\Longleftrightarrow\;\;
\Tpz = e V_0 \sin(y_0/\lambda)\, e^{-z_0/\lambda},
\label{eq:balance}
\end{equation}
the published balance condition, equivalent point by point to energy conservation along
the straight line: the potential climb equals the transverse drain, which is \emph{why}
the two fields must share $\lambda$. The transverse energy on every guiding-center path,
balanced or not, obeys $d\Tp/dz = -\Tp/\lambda$.

\subsection{The exact invariant and the off-balance theorem}
\label{sec:invariant}

Define $D \equiv \mu B(z) + q\phi(y,z)$. Along the drift flow, $dD/dt = 0$ exactly ---
$D$ is the guiding-center Hamiltonian, and the drift trajectories are its level curves;
the balanced line is the level set $D = 0$. The natural temptation is to evolve the
mismatch $\Delta\Tp$ itself along the curved trajectory; the conserved object is $D$,
and the transverse energy remains a function of $z$ alone wherever the electron is. This
dissolves the apparent tension between exact $\mu$-conservation and off-balance
dispersion: the dispersive displacement is driven by a conserved quantity, not by a
changing invariant.

For $\mu$ itself we assemble the standard adiabatic
canon~\cite{kruskal1962,northrop1963,littlejohn1983} into the statement the instrument
needs: for these fields (smooth, analytic, with $\wc \neq 0$ through the filter), the
guiding-center expansion furnishes an invariant series conserved to all orders in
$\varepsilon = \rho/\lambda$ along every trajectory in the plane, and off-balance
operation does not enter the expansion at gyro-frequency, so it cannot produce a secular
drift at any finite order. The all-orders statement attaches to the dressed invariant of
the expansion --- the series whose leading term is the bare moment $p_\perp^2/2mB$; the
bare moment carries bounded $O(\varepsilon)$ finite-orbit oscillations about it, and it
is the dressed invariant that the instrument's transport freezes. The true
non-conservation is beyond all orders~\cite{hastie1969}, with the adiabaticity exponent
\begin{equation}
\frac{\Delta\mu}{\mu} \sim
\exp\!\left[-\,C\,\varepsilon(z)^{-2}\right],
\qquad C = O(1),
\label{eq:leakage}
\end{equation}
a $1/\varepsilon^2$ law, not $1/\varepsilon$: the field changes on the drift transit
time, itself $1/\varepsilon$ slower than the gyro crossing. The factorized drift-time
estimate gives $C \approx 2/\coty$; its value is left to a dedicated measurement
(Sec.~\ref{sec:adiabaticlimit}). The exponent is parametrically enormous at the
published entrance, $C\,\varepsilon_0^{-2}$ with $\varepsilon_0^{-2} = 2.3\times10^4$,
and reaches order unity at the adiabatic limit (Sec.~\ref{sec:adiabaticlimit}):
parametrically, the limit and the onset of real index change are the same surface.

Quantum reading: conserving the invariant is conserving the Landau index $n$ (the
invariant \emph{is} the action). The ideal filter --- balanced or dispersive --- never
changes $n$; the rung spacing $\hbar\wc$ collapses beneath a frozen
index~\cite{dodonov2021,dodonov2023}. Radiation is the only de-excitation channel
(Sec.~\ref{sec:radiation}).

\subsection{The off-balance closed form}
\label{sec:offbalance}

\begin{figure}[tb]
\includegraphics[width=\columnwidth]{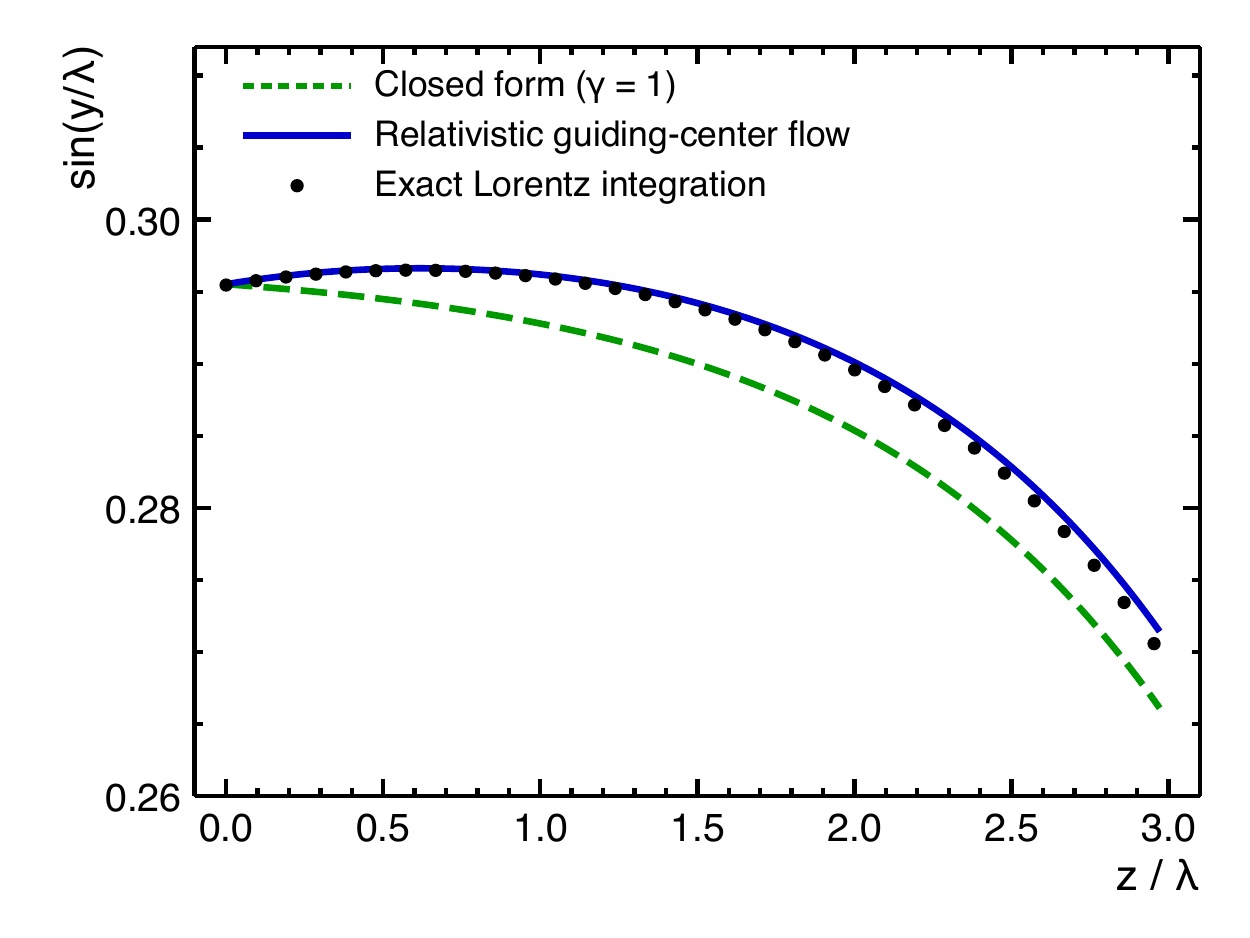}
\caption{Off-balance trajectory at $\Delta\Tp = +100\eV$. The closed form
[Eq.~(\ref{eq:corrected}), dashed] is the exact solution of the $\gamma = 1$ drift
equations; the relativistic guiding-center flow (solid) adds the bounded
drive-mismatch offset of Sec.~\ref{sec:relativistic} (the visible separation), and
the exact Lorentz integration (points) follows it. With that common-mode offset
cancelled between $\pm100\eV$ mismatch pairs, exact dynamics matches the closed form's
dispersion response to $1.6\times10^{-3}$ of signal.}
\label{fig:offbalance}
\end{figure}

For release at $(y_0, z_0 = 0)$ with mismatch
$\Delta\Tp \equiv \Tpz - eV_0\sin(y_0/\lambda)$, the drift equations integrate in closed
form to
\begin{equation}
\sin\!\bigl(y(z)/\lambda\bigr) = \sin(y_0/\lambda)
- \frac{\Delta\Tp}{eV_0}\bigl(e^{z/\lambda} - 1\bigr):
\label{eq:corrected}
\end{equation}
an energy mismatch is magnified exponentially into transverse displacement, and the
acceptance window closes as $e^{-z/\lambda}$. Exact relativistic
Lorentz-force integration confirms Eq.~(\ref{eq:corrected})
(Fig.~\ref{fig:offbalance}): with common-mode second-order drift corrections cancelled
between $\pm100\eV$ mismatch pairs, the exact dynamics follows the closed form to
$1.6\times10^{-3}$ of the mismatch signal over three decay lengths, and the measured
exponential growth factor matches $e^{z/\lambda}$ to $0.1\%$.

\subsection{Relativistic shapes}
\label{sec:relativistic}

\begin{figure*}[tb]
\includegraphics[width=0.49\textwidth]{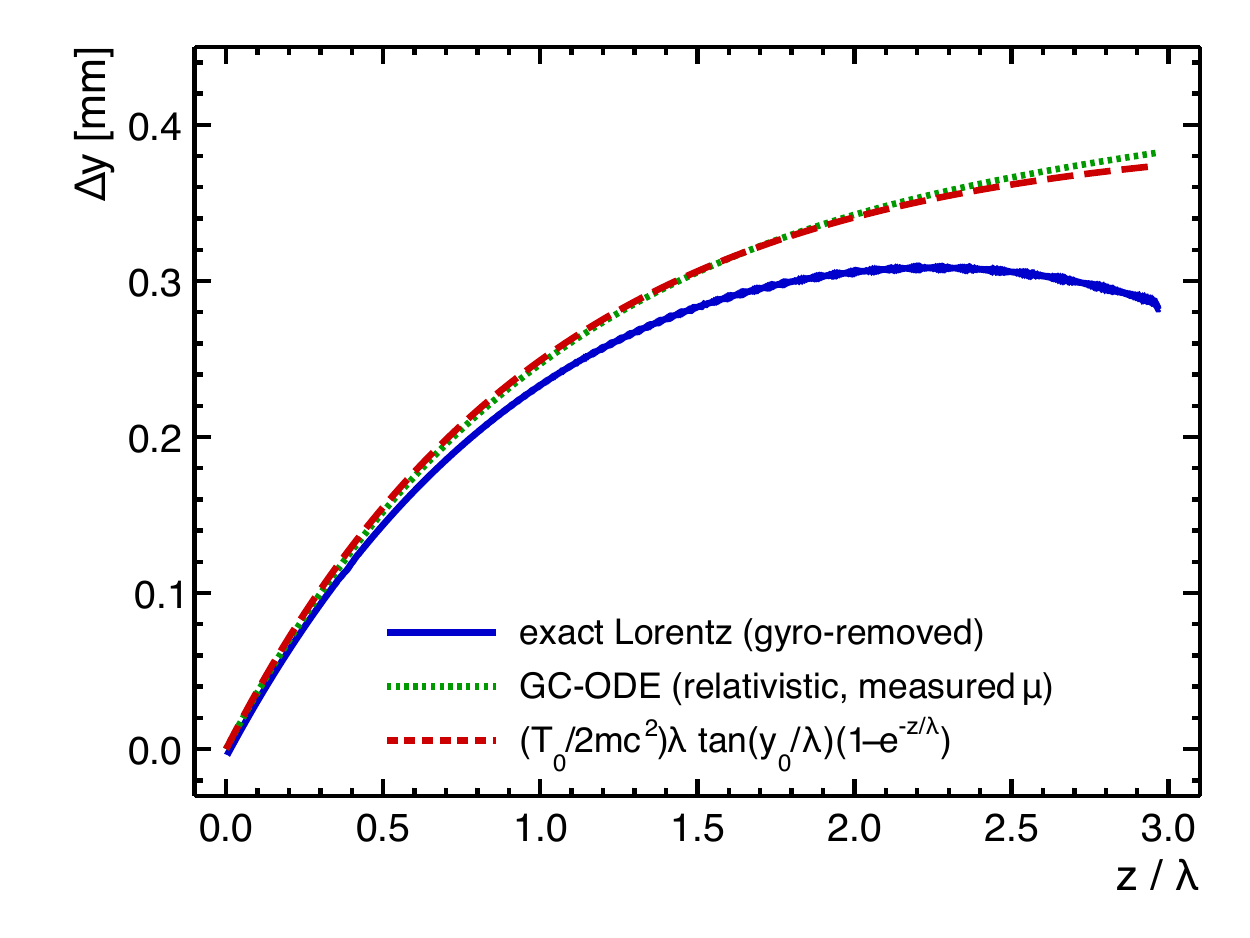}\hfill
\includegraphics[width=0.49\textwidth]{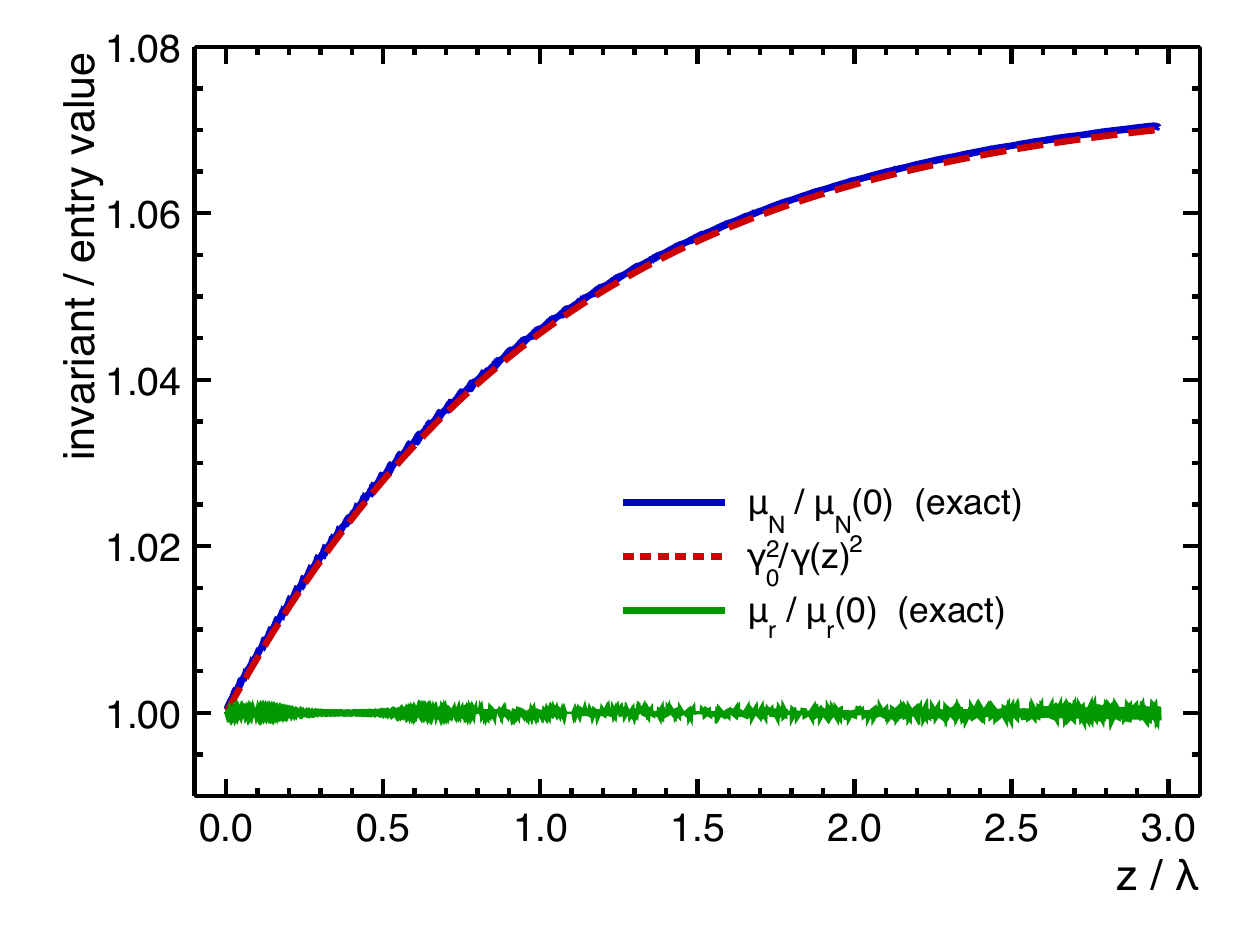}
\caption{Relativistic calibration shapes in the ideal fields, exact integration.
Left: the saturating transverse offset of a non-relativistically matched drive ---
exact dynamics, the relativistic guiding-center flow at the measured invariant, and the
closed formula of Eq.~(\ref{eq:offset}) (the latter pair agree to $0.3\%$); the exact
trajectory sits ${\approx}11\%$ below both, consistent in sign and size with the
second-order drift corrections not retained in Eq.~(\ref{eq:offset}). Right: the non-relativistic moment
$\mu_N = m v_\perp^2/2B$ rises by $\gamma_0^2 - 1 = 7.4\%$ (asymptotically) with the
predicted profile $\gamma_0^2/\gamma(z)^2$, while the relativistic invariant
$\mu_r = p_\perp^2/2mB$ stays flat: the published few-percent ``moment increase'' has
exactly the size and shape of this artifact.}
\label{fig:relativistic}
\end{figure*}

At $18.6\keV$ ($\gamma_0 = 1.0364$) the relativistic invariant is $p_\perp^2/2mB$; the
matched drive carries the factor
$\kappa_{\mathrm{rel}} \equiv (\gamma_0+1)/2\gamma_0 = 0.982$; and a voltage profile
programmed with the non-relativistic drain law is off balance by $1.8\%$ at entry,
decaying with the drain. The resulting transverse offset is \emph{bounded}: the
exponential amplification and the decaying forcing cancel into a saturating displacement
\begin{equation}
\Delta y(z) = \frac{T_0}{2 m_e c^2}\,\lambda \tan(y_0/\lambda)
\bigl(1 - e^{-z/\lambda}\bigr),
\label{eq:offset}
\end{equation}
about $0.4$~mm at the reference point --- the ``semi-relativistic deviation'' visible in
the published simulations~\cite{betti2019design}, now with a formula, and absorbable by
programming the relativistic drain law into the electrodes. Separately, the
\emph{non-relativistic} moment $m v_\perp^2/2B$ is not conserved: it rises by
$\gamma_0^2 - 1 = 7.4\%$ over the drain ($63\%$ of it within the first decay length)
while $p_\perp^2/2mB$ stays flat. Exact integration gives both behaviors --- a measured
rise of $7.03\%$ over three decay lengths against $7.00\%$ predicted, with the true
invariant flat to $1.3\times10^{-3}$ (Fig.~\ref{fig:relativistic}) --- and identifies
this artifact as the likely origin of the few-percent moment increase reported in the
published simulations~\cite{apponi2022implementation}. Apportioning the published
increase between this artifact and the implemented device's real entrance-field
transition would require that simulation's configuration and is not attempted here; the
entrance-kink contribution is bounded analytically at ${\sim}2\times10^{-5}$ and cannot
carry the few percent, while the relativistic artifact has exactly the right size and
profile.

\section{The exact quantum guiding-center theory}
\label{sec:quantum}

\subsection{Canonical structure and the squeeze}
\label{sec:squeeze}

The guiding-center coordinates of a charge in a perpendicular field do not
commute~\cite{maraner1997,andradesilva2026guiding}:
$[\hat Y, \hat Z] = i\lB^2(z)$ with $\lB = \sqrt{\hbar/eB}$. The flux-type momentum
$P \equiv -e B_0 \lambda\, e^{-z/\lambda}$ canonicalizes the drift plane
($[\hat y, \hat P] = i\hbar$), and the conserved $D$ becomes
\begin{equation}
D = P\, v_y(y), \qquad
v_y(y) = \frac{V_0}{B_0\lambda}\sin(y/\lambda) - \frac{\mu}{e\lambda},
\label{eq:Dcanonical}
\end{equation}
the product of the canonical momentum with the net vertical drift profile. Near balance,
with $\kappa \equiv v_z/\lambda$,
\begin{equation}
D \approx \kappa\, \delta y\, P,
\label{eq:squeezegen}
\end{equation}
the canonical squeeze (dilation) generator. Three operator identities follow for the
quadratic flow: $\delta\hat y(t) = \delta\hat y(0)\, e^{\kappa t}$,
$\hat P(t) = \hat P(0)\, e^{-\kappa t}$, $\delta\hat z(t) = \delta\hat z(0)$. The
$y$-spread of \emph{any} state grows with the full classical exponent; the $z$-spread is
frozen; no state narrows in $y$ under the drift flow. Exact integration tracks the
squeeze law to $6\times10^{-3}$ over $2.5$ decay lengths, with $z$-translates at fixed
invariant preserved within the $6\%$ extraction floor of the offset diagnostic.

\subsection{Exact solvability, criticality, and the quantized balance}
\label{sec:solvable}

Three structures are exactly solvable, and one is not:
\begin{itemize}
\item The full 3D microscopic problem does not separate (the drive breaks
$y$-translation; the off-plane field couples the bounce) --- stated plainly, since hope
to the contrary is natural. The off-plane (bounce) sector --- its drain in the
implemented geometry, and its role in entrance-state preparation --- lies beyond the
drift theory of this paper (Sec.~\ref{sec:meaning}).
\item The undriven slice ($V_0 = 0$, $x = 0$) is a Morse
problem~\cite{fakhri2010,handrich2005}: the exponential magnetic skeleton supports exact
drifting Landau states, spectrum
$E_{n,k} = \hbar\wc(z_k)\,(n+\tfrac12) - \hbar^2 (n+\tfrac12)^2/2m\lambda^2$, with the
fractional correction to the level spacing --- in the local sense $\partial E/\partial n$
--- exactly the classical ordering parameter
$\varepsilon_n^2/2$ (the anharmonic term itself is $(n+\tfrac12)\,\varepsilon_n^2/4$ of
the spacing), and the quantum gradient-$B$ drift reproducing the classical expression
with eigenvalues in place of energies.
\item The guiding-center drift sector of the \emph{generalized} (off-balance) filter is
exactly solvable for any profile: first order in $\hat P$, with closed-form
$\delta$-normalized eigenfunctions; the spectrum is continuous on the open level curves,
so no spurious drift quantization is imposed.
\item At the balance line the drift eigenfunctions become the scale-invariant critical
states of the dilation generator --- the balanced filter operates at the critical point
of a hyperbolic flow, with a natural log-periodic scale $\hbar\kappa \approx 2.5$~neV
that sets no floor but marks the idealized transmission structure.
\end{itemize}

The balance condition itself quantizes: with $\mu$ frozen at $(n+\tfrac12)\hbar e/m$,
exact balance selects one Landau level per drive voltage,
\begin{align}
e V_0 \sin(y_0/\lambda) &= (n+\tfrac12)\,\hbar\wcz,
\nonumber\\
\Delta V_0 = \frac{\hbar\wcz}{e\sin(y_0/\lambda)}
&\approx \frac{0.12\,\mathrm{mV}}{\sin(y_0/\lambda)} \;\text{at 1 T},
\label{eq:quantbalance}
\end{align}
the filter's version of the discrete flat-band sequence known from the condensed-matter
analogue of this configuration~\cite{huang2026flatband}. Observing it in the keV
instrument requires the selectivity of Sec.~\ref{sec:ent} pushed below one rung
(entrance squeezing by ${\sim}60$) together with part-per-billion voltage control; we
state it as structure, not as a near-term observable in the keV instrument.

\section{The entrance limit}
\label{sec:ent}

\subsection{Magnification is not resolution}
\label{sec:magnification}

One might hope the dispersive mode's resolution improves with the local magnetic length
at the sensor, $\lB(z_s) = \lB(0)\, e^{z_s/2\lambda}$, against a signal growing as
$e^{z_s/\lambda}$ --- a resolving power improving as $e^{z_s/2\lambda}$. It does not.
The arriving state is not a fresh coherent state of the local field; it is the squeezed,
propagated entrance state, wider than $\lB(z_s)$ by the exponentially large factor
$(\delta y_0/\lBz)\, e^{z_s/2\lambda}$. Because $D$ is conserved, no unitary drift
evolution can improve a measurement of $D$ beyond the information in the initial
state~\cite{clerk2010,degen2017sensing}: the filter is an ideal telescope viewing its
own entrance slit.

\subsection{The limit}
\label{sec:entformula}

The floor of the dispersive mode is therefore set at the entrance,
$\sigma_{\mathrm{ent}} = |\partial D/\partial y|_0\, \delta y_0$ expressed as an
entrance energy, i.e.
$\sigma_{\mathrm{ent}} = (\Tpz\coty/\lambda)\,\delta y_0$,
and for an unsqueezed (coherent-scale) entrance, $\delta y_0 = \lBz/\sqrt{2}$:
\begin{equation}
\boxed{\;
\sigma_{\mathrm{ent}} = \frac{\Tpz\coty}{\lambda}\,\frac{\lBz}{\sqrt 2}
\approx 4.8\meV\cdot\coty
\;}
\label{eq:ent}
\end{equation}
at $1$~T, $\lambda = 7$~cm, $18.6\keV$ --- length-independent, scaling as
$\Tpz\, B_0^{-1/2} \lambda^{-1} \coty$ --- about forty entry Landau quanta, and within
one order of magnitude of extrapolated electrode-voltage floors: at tesla-scale entry
the entrance limit dominates any sub-meV selection window unless the geometry or the
state is engineered. $\hbar$ enters once, through $\lBz$; the classical limit recovers
an unbounded selection.

Three floors, in strict hierarchy: (i) the gyro-orbit smear
$(eV_0\cos/\lambda)\,\rho_0\, e^{-z/2\lambda}$ --- the surviving half-exponent law ---
which dominates every realizable device ($\approx 0.8\eV\cdot\cot$ at the adiabatic
limit, matching the published exit-uncertainty band, as it should); (ii) the entrance
limit; (iii) the local magnetic length, smaller than the propagated spread by
$e^{-z/2\lambda}$, which is never operative. The smear is not a true floor: it is
coherent gyration, removable by detection that resolves the exit orbit --- and with
orbit-fitted arrival extraction the entrance limit is attained from exact dynamics at
technical entrance spreads.

One measured caveat governs the quantum-scale regime. The transit dresses the arriving
state at the \emph{second} harmonic of the entrance gyrophase, with entrance-referred
amplitude of order $\varepsilon_0\rho_0$ and measured $\varepsilon$-linear scaling; at
coherent-scale entrance spreads this residue survives the exit-orbit fit itself and
holds the landing readout a factor ${\approx}60$ above the entrance limit at the
reference geometry, with first-harmonic (nanosecond-timing) phase correction providing
no improvement. The response is a measured, coherent function of the entrance phase, so
it is removable in principle given the preparation's phase record; no exit-side-only
readout has been demonstrated to remove it, and the accessibility of the limit at the
coherent scale through this second-harmonic dressing is open, alongside the
correlated-entrance question below. Entrance squeezing by $s$ lowers the limit by $s$
until single-rung granularity ($\hbar\wcz$ at the release field) takes over; a
squeezed-optimum refinement $(\lBz/\lambda)\sqrt{eV_0\cos\cdot\Delta\Tp}$ applies below
${\sim}$eV mismatches.

The balanced filter's selection sharpness obeys the same entrance theorem --- the
balanced mode is the dispersive mode operated at $D = 0$ with the acceptance edge as the
sensor --- so one formula bounds both operating modes.

\emph{Scope of the entrance-limit theorem.} The derivation factorizes the gyration from the
guiding-center sector, and the one identified threat is their $O(\varepsilon)$ coupling:
an entrance state correlating the two could, in principle, let gyro-side measurements
assist the energy estimate below the limit. The class the theorem addresses is the
unchirped minimum-uncertainty displacement family --- Gaussian guiding-center states of
width $\delta y_0 = s\,\lBz/\sqrt{2}$, displaced by the energy being encoded, in product
with a gyration state carrying no dependence on that energy. For this class the
protection is twofold. First, the transit is a unitary that does not depend on the
conserved energy, so it cannot change the quantum Fisher information the state carries
about that energy, at any order in $\varepsilon$; any non-postselected detection
extracts at most that information per incident electron, and postselection concentrates
information into detected events only by paying success probability, which re-enters the
error budget as statistics. Second, for preparation families symmetric under gyrophase
rotation --- the symmetry must hold for the complete preparation, retained records
included --- the odd parity of the coupling under gyro reflection forbids a
first-order-in-$\varepsilon$ change in that same Fisher information; corrections enter
at second order in $\varepsilon$, whose coefficient we do not bound here. The
first-order secular coupling itself cancels --- the same cancellation that protects
$\mu$, consistent with the quantal guiding-center normal
form~\cite{maraner1996,maraner1997} and measured directly in exact-dynamics ensembles
(the first-harmonic secular landing response nulls at the $6\times10^{-4}$ relative
level). What the theorem does not cover is entrance states that
\emph{do} carry gyro--drift correlation, and such states are not exotic: ordinary slit
collimation localizes the particle rather than its guiding center, which correlates the
guiding center with the gyrophase --- a preparation for which neither a prohibition nor
a gain is established, in either direction. The accessibility theory of correlated
entrances --- what a joint position--phase readout extracts, and what preparation it
requires --- is open, and it is the one place the entrance limit's coefficient could in
principle be revised; within the displacement family, the squeezing factor $s$ is the
entire freedom, and it is already included in the design space of Sec.~\ref{sec:optimum}.

\section{The radiation floor}
\label{sec:radiation}

\subsection{The power law and the mean radiated energy}
\label{sec:erad}

Spontaneous cyclotron emission~\cite{peil1999} is the ideal filter's only de-excitation
channel. For circular gyromotion the relativistic emission power takes the exact
two-line form
\begin{equation}
P = \frac{e^4 B^2 p_\perp^2}{6\pi\epsilon_0 m^4 c^3} = P_0\, e^{-3z/\lambda},
\label{eq:power}
\end{equation}
with no residual $\gamma$ once written in $p_\perp$, and
$P_0 = 7.34\keV/\mathrm{s}$ at $1$~T and $18.6\keV$; the exponential profile follows
from conservation of the moment at leading order (two powers of $B$ from the field, one
from $p_\perp^2$), in the circular-gyromotion model carrying measured $O(\varepsilon)$
finite-orbit corrections --- the emission is concentrated in the first $\lambda/3$.
Integrated along the transit at the balanced drift speed,
\begin{equation}
\boxed{\;
E_{\mathrm{rad}} = \frac{e^5 B_0^3 \lambda^2 \gamma_0}{9\pi\epsilon_0 m^3 c^3 \coty}
\;}
\label{eq:erad}
\end{equation}
--- \emph{independent of the electron's energy}: the balance condition ties the drive
voltage, hence the transit time, to the energy being filtered, and the energy cancels.
The invariance is non-relativistic and compares energy-retuned balanced settings: the
$\gamma_0$ prefactor carries the only energy dependence
($\partial E_{\mathrm{rad}}/\partial \Tpz = E_{\mathrm{rad}}/\gamma_0 m_e c^2 \approx
1.2\times10^{-9}$ at $1$~T, $\lambda = 7$~cm, $\cot = 1$), while within one instrument
at fixed drive $E_{\mathrm{rad}} \propto \Tpz$ --- uniform across any endpoint-class
window to parts in $10^3$. The constant-drift-speed quadrature behind Eq.~(\ref{eq:erad})
holds at the per-cent level along real trajectories; exact integration reproduces the
closed form at the $2\%$ level, with both identified corrections (the drive law; an
$O(\varepsilon)$ radiative-drift term) carrying their expected signs. Dimensionlessly,
with $\lambda_C \equiv \hbar/m_e c$ the reduced Compton wavelength and $\alpha$ the
fine-structure constant,
\begin{equation}
\frac{E_{\mathrm{rad}}}{\hbar\wcz} = \frac{4}{9}\,\alpha\,\gamma_0\,
\Bigl(\frac{\lambda}{\lambda_C}\Bigr)^{\!2}
\Bigl(\frac{\hbar\wcz}{m_e c^2}\Bigr)^{\!2} \frac{1}{\coty}:
\label{eq:rungs}
\end{equation}
about $6$ entry quanta ($0.66\meV$) at $1$~T and $0.65\eV$ at $10$~T, both at
$\cot = 1$, scaling as $B_0^3\lambda^2$.

\subsection{The fluctuation, exactly, and its mode-independence}
\label{sec:fluctuation}

The mean radiated energy is deterministic and correctable --- by the invariance, without
even knowing the electron's energy. The floor is the fluctuation. Emission is
compound-Poisson shot noise in which each photon carries the \emph{local} quantum
$\hbar\omega(z) = \hbar e B(z)/\gamma(z) m$, so the variance is the first moment of the
photon energy against the radiated energy:
\begin{equation}
\boxed{\;
\sigma^2_{\mathrm{rad}} = \int \hbar\omega\, dE_{\mathrm{rad}}
= \frac{3}{4}\, \hbar\wcz\, E_{\mathrm{rad}}\, c_\gamma,
\;}
\label{eq:variance}
\end{equation}
the $3/4$ from the $e^{-3z/\lambda}$ power profile against
$\hbar\omega \propto e^{-z/\lambda}$ (assigning every photon the full entry quantum
overstates $\sigma$ by $\sqrt{4/3}$), and $c_\gamma = 0.9717$ at $18.6\keV$ the exact $\gamma(z)$ quadrature.
At $\cot = 1$, $\lambda = 7$~cm: $\sigma_{\mathrm{rad}} = 0.235\meV$ at $1$~T, rising as
$B_0^2$ to $23.5\meV$ at $10$~T.

\begin{figure*}[tb]
\includegraphics[width=0.49\textwidth]{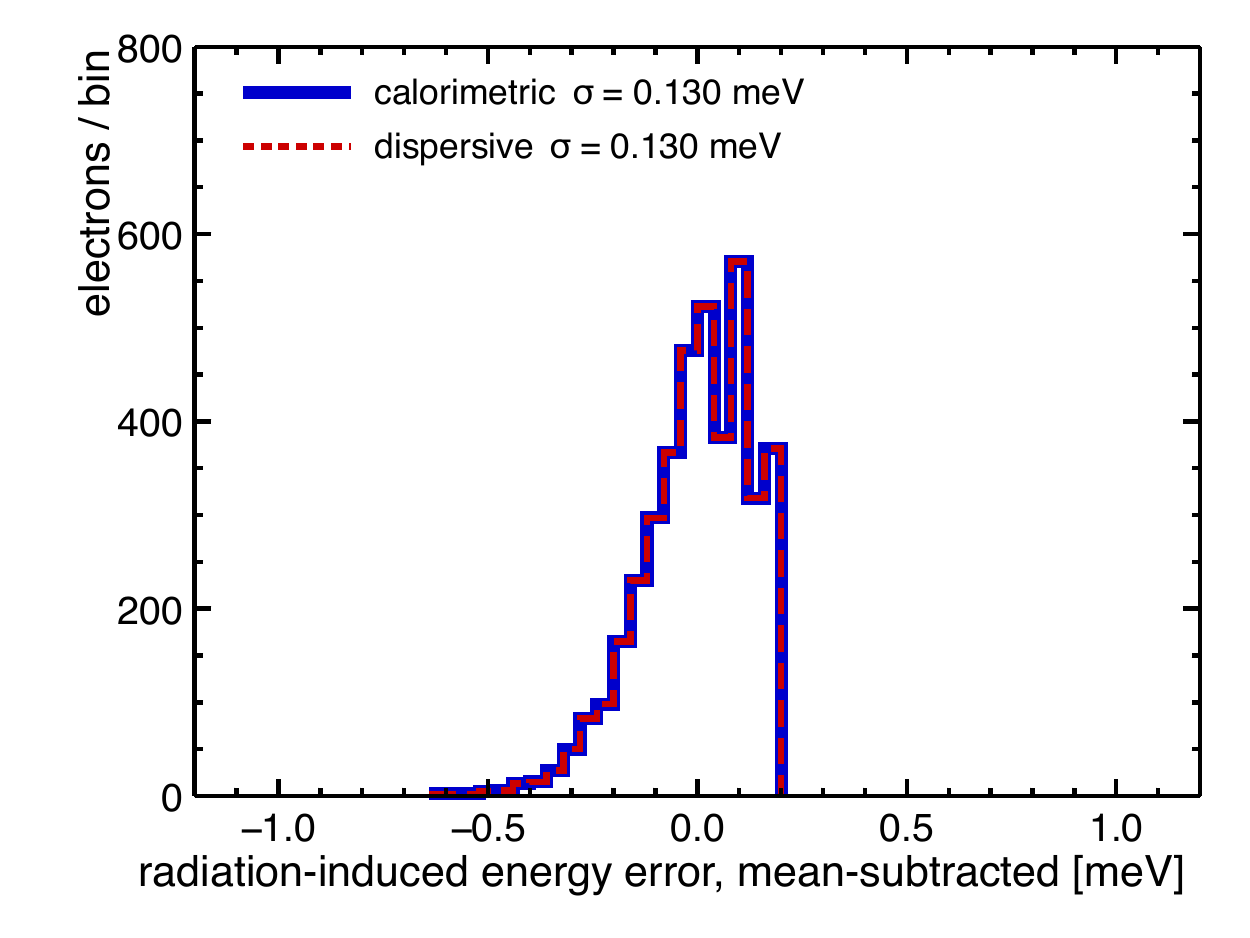}\hfill
\includegraphics[width=0.49\textwidth]{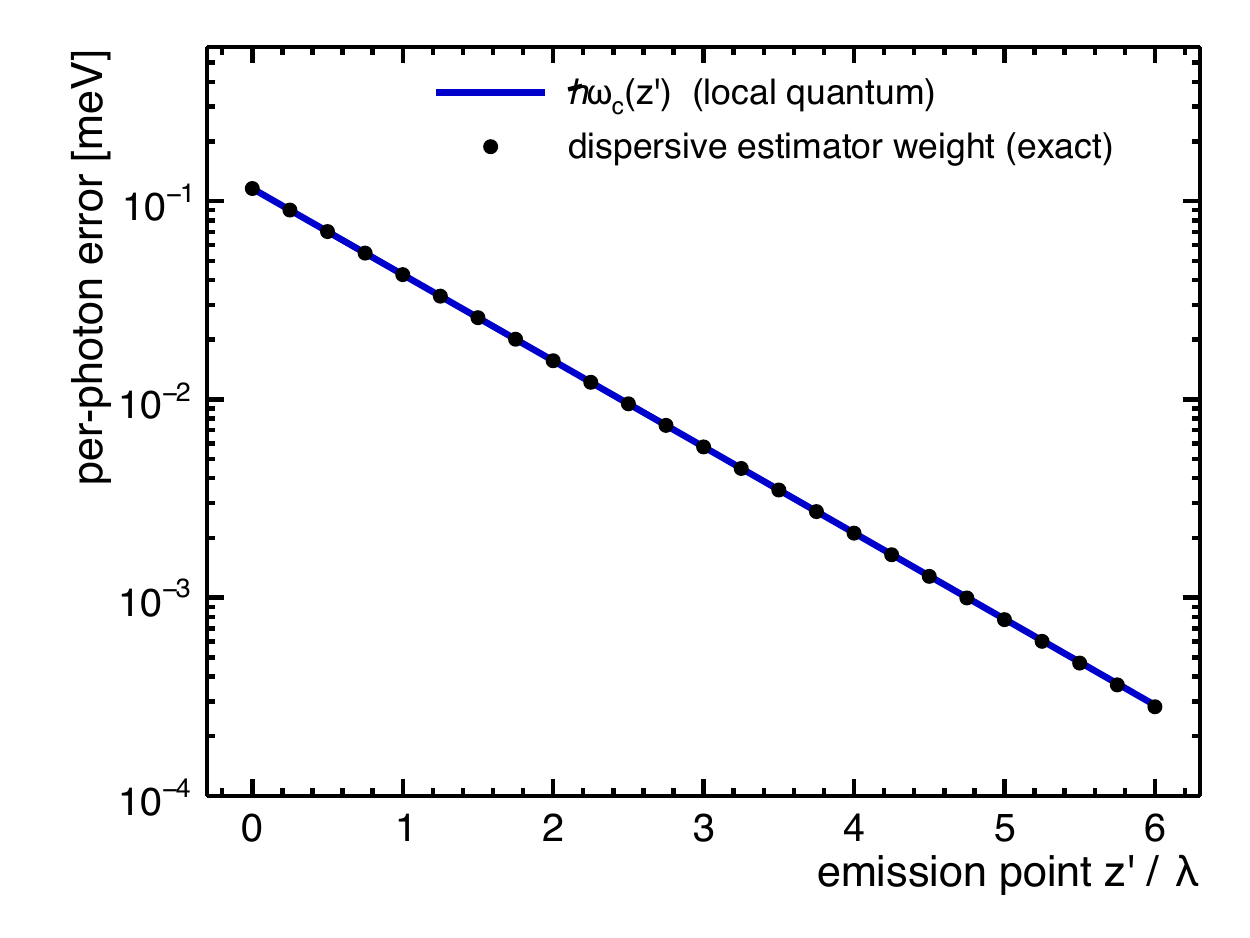}
\caption{The radiation floor is readout-mode-independent. Left: Monte Carlo over 4000
electrons with photon-by-photon emission at $1$~T, $y_0/\lambda = 0.3$, in the
fundamental-frequency $\gamma = 1$ model with the finite-exit weights of
Eq.~(\ref{eq:weight}) --- the calorimetric inferred-energy error
$-(E_{\mathrm{rad}} - \langle E_{\mathrm{rad}}\rangle)$ and the dispersive
inferred-energy error share one distribution. Right: the reason --- the landing-plane
weight of a one-rung step at $z'$ equals the local quantum $\hbar\wc(z')$ up to the
finite-exit end factor of Eq.~(\ref{eq:weight}) (points: exact single-photon weights;
curve: the local quantum).}
\label{fig:mcmodes}
\end{figure*}

The floor is independent of how the electron is read out (Fig.~\ref{fig:mcmodes}). In
the calorimetric mode the reading misses exactly the photon energies. In the dispersive
mode a one-rung step $\delta\mu = \hbar e/m$ at $z'$, propagated through the
landing-position readout, carries the filter's remaining exponential weight: it
shifts the inferred entrance energy by
\begin{equation}
\delta T_{\mathrm{est}}(z') =
-\,\hbar\wcz\,
\frac{w'/\gamma' \,-\, w_s/\gamma_s}{\,1 - \gamma_0\, w_s/\gamma_s\,},
\qquad w \equiv e^{-z/\lambda},
\label{eq:weight}
\end{equation}
with $w'$, $\gamma'$ evaluated at emission and $w_s$, $\gamma_s$ at the sensor --- for a
deep exit $z_s \gg z'$ this is precisely the $\gamma$-dressed local quantum
$-\hbar\wc(z')$, the same energy the calorimeter misses (a tempting ``$\sqrt{2}$ between
the modes,'' from reading the exit moment with uniform weight, is thereby ruled out).
The equality is a linear-response statement carrying the finite-exit factor of
Eq.~(\ref{eq:weight}); one number serves both readouts.

Two corrections carry design content. \emph{Thermal:} at bore temperature $T$ the
variance ratio in the fundamental-frequency, non-relativistic model is
\begin{equation}
R = 1 + \frac{8}{x^4}\int_0^{x} \frac{w^3\, dw}{e^w - 1},
\qquad x = \frac{\hbar\wcz}{k T},
\label{eq:thermal}
\end{equation}
a quartic, Stefan--Boltzmann-like turn-on $(8\pi^4/15)(kT/\hbar\wcz)^4$ for cold bores,
crossing to the Rayleigh--Jeans $(8/3)(kT/\hbar\wcz)$ far above the ground-state
crossover temperature $\hbar\wcz/k_B = 1.34\,\mathrm{K}\cdot B_0[\mathrm{T}]$;
suppression requires $kT \ll \hbar\wcz$ --- at $1$~T a $650$~mK bore already costs
$+58\%$ in variance (the exact $\gamma(z)$-retained quadrature; Eq.~(\ref{eq:thermal})
itself gives $+55\%$), while at the optimum fields of Sec.~\ref{sec:optimum} a $100$~mK
bore adds nothing (the sub-kelvin, quantum-cyclotron
regime~\cite{peil1999,hanneke2008}). \emph{Spectral:} at
$\beta_\perp = 0.26$ the harmonic content is not negligible --- the second harmonic
carries $(12/5)\beta^2 + O(\beta^4) = 0.16$ of the fundamental's power at entry ($0.13$
of the total), each harmonic-$n$ photon stepping the Landau index by $n$ with energy
$n\hbar\omega$ --- and the harmonic-summed variance (circular-orbit Schott sum with the
$\gamma$-retained drain weighting) is $1.145\times$ the fundamental-only value ($\sigma$
$\times 1.070$), front-loaded and scaling-preserving. The mean radiated energy is untouched --- the
emitted power already contains every harmonic; the spectrum moves only the fluctuation.

Any upstream dwell at full field (a frequency
tag~\cite{monreal2009,asner2015project8}; a transport delay) adds
\begin{equation}
\sigma^2 = (1-\eta_{\mathrm{eff}})\,(1 + 2\bar n)\, \hbar\omega\, P\, t_{\mathrm{dwell}},
\label{eq:dwell}
\end{equation}
with $\eta_{\mathrm{eff}}$ the photon-event detection efficiency (emission and
absorption separately in general); the breakeven against the transit floor is
$t^* = \lambda/4v_z = e B_0 \lambda^2/(4\kappa_{\mathrm{rel}}\Tpz\cot) \approx
67\,\mathrm{ns}\cdot(\lambda/7\,\mathrm{cm})^2/\cot$ at $1$~T and $18.6\keV$ (the
balanced speed itself scales as $\lambda^{-1}$) --- the cold, unobserved-photon design
equation for radiation-tagged operation.

\section{The optimal entrance field}
\label{sec:optimum}

\begin{figure}[tb]
\includegraphics[width=\columnwidth]{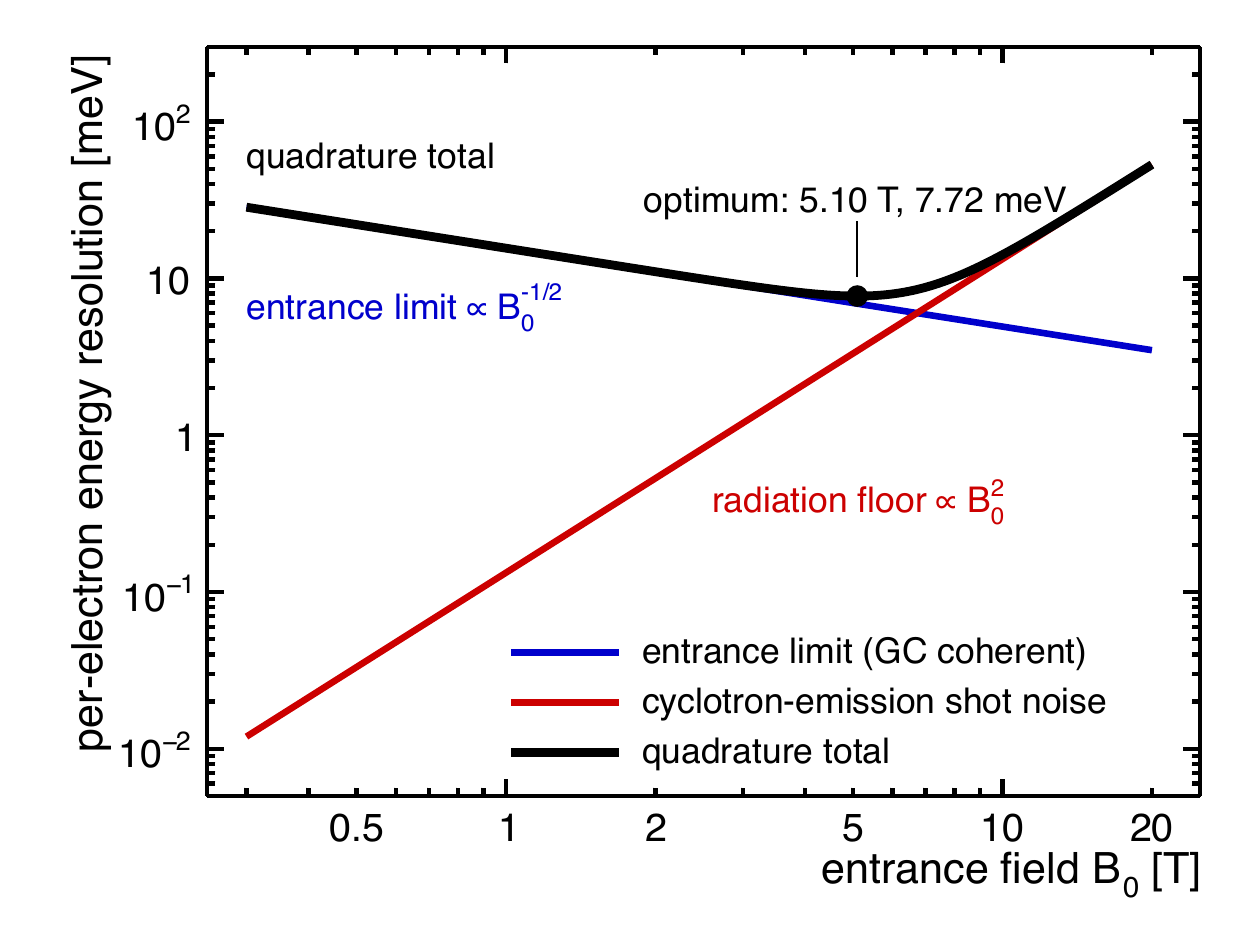}
\caption{The two quantum floors of the dispersive drift filter against entrance field,
at $18.6\keV$, $\lambda = 7$~cm, $y_0/\lambda = 0.3$, coherent entrance, cold bore, in
the fundamental-frequency ($c_\gamma \to 1$) model of Eq.~(\ref{eq:optimum}). The
entrance limit falls as $B_0^{-1/2}$, the radiation floor rises as $B_0^{2}$, and the
quadrature total has the closed-form interior minimum of Eq.~(\ref{eq:optimum}) with its
universal $4{:}1$ variance partition between entrance limit and radiation floor;
$B_{\mathrm{opt}} \propto \cot^{3/5}$ moves it to $2.5$~T at the slow-drive edge
$\cot = 1$.}
\label{fig:optimum}
\end{figure}

The dispersive instrument carries two per-electron $\hbar$-floors with opposite field
scalings:
\begin{align}
\sigma^2_{\mathrm{ent}} &= \frac{a}{B_0}, &
a &= \frac{\hbar\, \Tpz^2 \cot^2}{2 e \lambda^2};
\nonumber\\
\sigma^2_{\mathrm{rad}} &= b\, B_0^4, &
b &= \frac{\hbar e^6 \lambda^2 \gamma_0}{12\pi\epsilon_0 m^4 c^3 \cot}.
\label{eq:ab}
\end{align}
The quadrature total is strictly convex with a unique interior minimum
(Fig.~\ref{fig:optimum}):
\begin{equation}
\boxed{\;
\begin{aligned}
B_{\mathrm{opt}} &= \left[
\frac{3\pi\epsilon_0 m^4 c^3\, \Tpz^2 \cot^3(y_0/\lambda)}{2 \gamma_0 e^7 \lambda^4}
\right]^{1/5},
\\
\sigma_{\min} &= \sqrt{\tfrac{5}{4}}\; \sigma_{\mathrm{ent}}(B_{\mathrm{opt}}),
\end{aligned}
\;}
\label{eq:optimum}
\end{equation}
with the radiation variance exactly one quarter of the entrance-limit variance at the
optimum ---
one fifth of the total, a universal partition set by the exponents $-1$ and $+4$ alone
(the two floors themselves cross higher, at $4^{1/5}B_{\mathrm{opt}}$). At $18.6\keV$,
$\lambda = 7$~cm, in the coherent-entrance, fundamental-frequency ($c_\gamma \to 1$)
closed form: $B_{\mathrm{opt}} = 2.52$~T with $\sigma_{\min} = 3.39\meV$ at
$\cot = 1$; $5.10$~T with $7.72\meV$ at the thesis test geometry $y_0/\lambda = 0.3$
($\cot = 3.23$); dressed by the harmonic spectrum and the exact $\gamma(z)$ quadrature
together, ${\approx}2\%$ lower field and ${\approx}1\%$ higher floor.
Table~\ref{tab:floors} collects the floors and scales. A tesla-scale optimum across the
plausible geometry band --- the instrument's natural operating point is within reach of
ordinary superconducting magnets, and \emph{more field is not better}.

\begin{table}[tb]
\caption{\label{tab:floors}Quantum floors and scales of the dispersive drift filter at
the reference configuration ($18.6\keV$, $\lambda = 7$~cm, $\coty = 1$), for a
coherent-scale entrance ($\Delta Y_0 = \lBz/\sqrt{2}$), in the fundamental-frequency
($c_\gamma \to 1$) closed forms; $B_{\mathrm{opt}} = 2.52$~T at this configuration. The
gyro-orbit smear --- $0.8\eV\cdot\coty$ at the adiabatic limit, coherent gyration
removable by orbit-resolving detection (Sec.~\ref{sec:ent}) --- dominates every
realizable device and is not a quantum floor.}
\begin{ruledtabular}
\begin{tabular}{lccc}
 & Scaling & $B_0 = 1$~T & $B_0 = B_{\mathrm{opt}}$ \\
\hline
Entry Landau spacing $\hbar\wcz$ & $B_0$ & $0.116\meV$ & $0.292\meV$ \\
Entrance limit $\sigma_{\mathrm{ent}}$ & $B_0^{-1/2}$ & $4.8\meV$ & $3.0\meV$ \\
Mean radiated energy $E_{\mathrm{rad}}$ & $B_0^{3}$ & $0.66\meV$ & $10.5\meV$ \\
Radiation floor $\sigma_{\mathrm{rad}}$ & $B_0^{2}$ & $0.235\meV$ & $1.5\meV$ \\
Quantum floor (quadrature) & --- & $4.8\meV$ & $3.39\meV$ \\
\end{tabular}
\end{ruledtabular}
\end{table}

The design space, exponents exact:
\begin{align}
B_{\mathrm{opt}} &\propto
\Tpz^{2/5}\, \lambda^{-4/5}\, \cot^{3/5} \gamma_0^{-1/5} F_P^{-1/5}\, s^{2/5},
\nonumber\\
\sigma_{\min} &\propto
\Tpz^{4/5}\, \lambda^{-3/5}\, \cot^{7/10} \gamma_0^{1/10} F_P^{1/10}\, s^{4/5},
\label{eq:surface}
\end{align}
with $s \le 1$ the entrance-squeezing factor and $F_P \le 1$ a Purcell factor on the
emission environment. Readings: slowing the forward drive ($\cot \to 0$; the
$\pi/4$--$\pi/2$ region is geometrically accessible, and a slowed-drift mode has been
demonstrated~\cite{farino2025slowed}) lowers floor and field together; doubling
$\lambda$ improves the floor by only $2^{3/5} = 1.5$ --- the exponential went into
selectivity, not into these floors; squeezing by ten~\cite{caves1981} gains $\times6.3$
in floor while \emph{lowering} the optimal field to $0.40$ of its unsqueezed value ---
the squeezed instrument is sharper at lower field; a bore below the crossover
temperature $\hbar\wcz/k_B$ leaves the optimum unmoved (the quartic correction is
field-independent at that order), and a warm bore is never advantageous.

One scope statement completes the theorem: the optimum is a property of the dispersive
(selectivity) instrument class. A pure calorimetric record has no
entrance-position limit --- position uncertainty of any origin affects \emph{which}
electrons arrive, never the record --- and hence no optimum: its quantum noise level is
the radiation floor alone, lower entrance field is strictly better, and what limits
calorimetric energy reconstruction here is unrecorded radiation, not the entrance limit.

\section{The adiabatic limit on length}
\label{sec:adiabaticlimit}

All orderings of the guiding-center expansion collapse onto one parameter,
$\varepsilon(z) = \rho(z)/\lambda = \varepsilon_0\, e^{z/2\lambda}$ --- the exponential
profile is the unique profile that makes the field-variation ratio uniform --- with the
kinematic ratio $v_D/v_\perp = (\cot/2)\,\varepsilon$ proportional to it. Adiabatic
transport ends at $\varepsilon_c = O(1)$:
\begin{equation}
R_{\max} = e^{z_{\max}/\lambda} = \varepsilon_c^2\,
\Bigl(\frac{\lambda}{\rho_0}\Bigr)^{\!2},
\label{eq:Rmax}
\end{equation}
the maximum reduction ratio: $(\lambda/\rho_0)^2 \approx 2.3\times10^4$ at $1$~T,
$\lambda = 7$~cm, with the published device gap-limited at $\varepsilon_c \approx 0.25$
(reproducing its ${\sim}10^3$ field ratio and sub-meter length) --- a documented
geometric limit, sitting beneath the transport (leakage) cutoff of
Eq.~(\ref{eq:leakage}). The limit is a transport failure, not an energy
cliff~\cite{evers1998,dahlem2001} --- the rungs at exit are ${\sim}0.1\,\mu$eV --- and
the quantum $R_{\max}$ replaces $\rho_0$ by $\lBz$: larger by exactly $2n+1$ (an
identity exact relativistically, $\rho_n^2 = (2n+1)\lB^2$), the limit available to a
ground-state electron, which the keV instrument, its index frozen, never reaches.
Staging --- re-raising the field to re-compress the orbit at the cost of re-raising the
energy --- multiplies the attainable ratio geometrically at linear growth in length.

The $O(1)$ constant $C$ in $\Delta\mu/\mu \sim A\exp(-C/\varepsilon^2)$, which would fix
$\varepsilon_c$ quantitatively, is deferred: measuring it honestly requires a
volume-preserving integrator, because near the limit an adaptive integrator's secular
drift masquerades as leakage at exactly the exponents in question. $\varepsilon_c$ is
therefore carried as a parameter. No principal result of this paper depends on its
value; the
instrument's ultimate length and staging budgets depend on little else, and its
measurement is taken up with those budgets in the companion paper.

\section{Discussion of the limits}
\label{sec:meaning}

The exponential gain of adiabatic drift filtering is magnification, not resolving power:
for the dispersive (selectivity) instrument class, operated on coherent-scale entrances,
the quantum limits are two: the entrance limit, set once and for all at the entrance
plane, and the radiation floor on the energy balance itself --- meeting at a tesla-scale
optimal field with a universal variance partition between them --- while a pure
calorimetric record, which has no entrance limit, is bounded by the radiation floor
alone. Neither limit is close to current operation (the realizable device is
gyro-smear-limited, two to three orders of magnitude above the entrance limit), but both
are close to proposed operating targets: sub-meV selection windows at tesla entry fields
are entrance-limited, and the handles on the floors --- drive angle, decay length,
entrance squeezing, emission environment --- are design choices, not constants of
nature. The consequences are developed in a companion paper:
the mechanism's reach when staging, whole-window capture, and radiation-as-counter are
taken to their limits, with relic-neutrino detection as the target that sets every
requirement; the preparation of squeezed entrance states, in which the off-plane
(bounce) sector enters as a resource rather than a nuisance; the leakage constant that
sets the adiabatic limit; and the low-index realization of this theory in patterned-field
two-dimensional electron
systems~\cite{nogaret2010inhomogeneous,zhang2005graphene,huang2026flatband}, where the
quantized balance condition and the entrance limit become bench-top transport
measurements.

\bibliography{refs}

\end{document}